\documentclass[conference]{IEEEtran}

\usepackage{booktabs} 

\usepackage{setspace}
\usepackage{placeins}
\usepackage{afterpage}
\usepackage[utf8]{inputenc}
\usepackage{graphicx}
\usepackage{balance}
\usepackage{amsthm}
\usepackage{color}
\usepackage{caption}
\usepackage{amsmath}
\usepackage{tabu}
\usepackage{array}
\usepackage{booktabs}
\usepackage{threeparttable}
\usepackage{relsize}
\usepackage{physics}

\usepackage{algorithm}
\usepackage[noend]{algpseudocode}

\newlength{\maxwidth}
\newcommand{\algalign}[2]
{\makebox[\maxwidth][r]{$#1{}$}${}#2$}

\usepackage{float}
\usepackage{stfloats}
\usepackage{lipsum}
\usepackage[english]{babel}

\usepackage{comment}
\usepackage{url}

\usepackage[T1]{fontenc}
\usepackage{mwe}    
\usepackage{subfig}
\usepackage{amssymb}
\let\oldnl\nl
\newcommand{\nonl}{\renewcommand{\nl}{\let\nl\oldnl}}

\newtheorem{problem}{Problem}

\theoremstyle{definition}
\newtheorem{defn}{Definition}

\def\Equal{\texttt{=}}

\begin{document}

\title{ILB: Graph Neural Network Enabled Emergency Demand Response Program For Electricity}

\author{\IEEEauthorblockN{Sina Shaham}
\IEEEauthorblockA{\textit{University of Southern California} \\
Los Angeles, USA \\
sshaham@usc.edu}
\and
\IEEEauthorblockN{Bhaskar Krishnamachari}
\IEEEauthorblockA{\textit{University of Southern California} \\
Los Angeles, USA \\
bkrishna@usc.edu}
\and
\IEEEauthorblockN{Matthew Kahn}
\IEEEauthorblockA{\textit{University of Southern California} \\
Los Angeles, USA \\
kahnme@usc.edu}
}

\maketitle

\begin{abstract}
Demand Response (DR) programs have become a crucial component of smart electricity grids as they shift the flexibility of electricity consumption from supply to demand in response to the ever-growing demand for electricity. In particular, in times of crisis, an emergency DR program is required to manage unexpected spikes in energy demand. In this paper, we propose the Incentive-Driven Load Balancer (ILB), a program designed to efficiently manage demand and response during crisis situations. By offering incentives to flexible households likely to reduce demand, the ILB facilitates effective demand reduction and prepares them for unexpected events. To enable ILB, we introduce a two-step machine learning-based framework for participant selection, which employs a graph-based approach to identify households capable of easily adjusting their electricity consumption. This framework utilizes two Graph Neural Networks (GNNs): one for pattern recognition and another for household selection. Through extensive experiments on household-level electricity consumption in California, Michigan, and Texas, we demonstrate the ILB program's significant effectiveness in supporting communities during emergencies.
\end{abstract}


\section{Introduction}


Demand Response (DR) programs have emerged as an integral component of smart electricity grids. By shifting the flexibility of power consumption from the supply side to the demand side, these programs manage the continually increasing demand for electricity. Such programs incentivize households to reduce their electricity usage, thereby enhancing the agility and responsiveness of the network. The benefits of DR programs are manifold, including, but not limited to, preventing the need for new power plants, lowering electricity costs by avoiding high-priced energy purchases, enhancing grid reliability to prevent blackouts, and mitigating environmental damage by reducing the consumption of fossil fuels. An example of such a program is Time-of-Use pricing (TOU), where the cost of electricity varies based on the time of day, typically increasing during peak periods~\cite{torriti2012}.


A more diligent and critical type of DR programs is designed for times of crisis and is referred to as an emergency DR program. Such programs are developed to manage unexpected spikes in energy demand during emergency situations. Unfortunately, most of the existing emergency DR programs are only activated after the accident and without prior provisions made for households, such as during extreme weather events, unexpected equipment failures, or other emergencies. Hence, they mostly work as mediators of the damage rather than the prevention mechanism, which can have severe consequences for households and businesses. 
For example, during a heat wave in California in August 2020, the California Independent System Operator (CAISO) called for an emergency DR program to reduce electricity usage and prevent blackouts~\cite{caiso2020}. This included asking businesses to reduce their energy consumption during peak demand periods, as well as calling on residential customers to save energy during the hottest parts of the day. In another unfortunate incident during a winter storm in Texas in February 2021, the Electric Reliability Council of Texas (ERCOT) called for an emergency DR program to reduce electricity usage and prevent blackouts~\cite{busby2021cascading}. This included asking businesses and residents to conserve energy by reducing heating and hot water usage, as well as turning off non-essential appliances and electronics.

The earlier examples illustrate the requirement for more advanced emergency DR programs that can greatly ease the pressure on the power grid during crises and avoid losses to communities. Nevertheless, there exist several obstacles that have impeded the development of efficient DR programs, including the need for real-time communication and monitoring of energy usage, precise prediction of energy demand, and ensuring sufficient resources are available to meet the heightened demand during emergencies. 
Thankfully, recent progress in Machine Learning (ML) combined with the increased adoption of Advanced Metering Infrastructure (AMI) in households has cleared the way for the implementation of new DR programs~\cite{eia}. 
AMI meters are capable of measuring and recording electricity usage at least once per hour and transmitting this information to both the utility and the customer at least once per day. The range of AMI installations varies from basic hourly interval meters to real-time meters with built-in two-way communication capabilities that allow for instantaneous data recording and transmission.

Considering recent advancements, we propose an incentive-based program accompanied by a novel ML-driven approach to address the existing challenges and help to balance supply and demand in times of crisis. In particular, we have the following contributions:

\begin{itemize}
    \item We propose an innovative incentive-based strategy designed to manage demand and response effectively during critical scenarios. This approach, termed the Incentive-Driven Load Balancer (ILB), strategically identifies households with the flexibility to adjust their electricity consumption and then operates on a meticulously designed incentive system, motivating these families to moderate their energy usage in emergency situations.
    \item We develop an ML-driven two-step framework for the efficient selection of participants in the program. The framework involves two GNNs: a pattern recognition GNN and a household selection GNN. The pattern recognition GNN is a time-series forecasting dynamic graph with the goal of revealing similarities among users based on attention mechanism, taking into account socio-economic factors influencing the elasticity of demand. Meanwhile, the household selection GNN models communities and aids in selecting suitable households for the program based on the pattern recognition GNN's output. Our framework is thoughtfully developed to factor in geo-spatial neighborhoods.
    \item We have devised and publicly shared a dataset based on the factors that influence the elasticity of electricity demand for data mining objectives. This dataset is greatly empowered by a sample from the synthetically generated household-level data in~\cite{thorve2023high}, which offers a digital twin of residential energy consumption. 
    We further enrich this dataset by integrating education and awareness factors from high schools~\cite{edgap}, college education data from the Census~\cite{education}, median household income and unemployment statistics from the US Department of Agriculture~\cite{usda}, and climate data~\cite{NCEI}.
    \item We perform an extensive number of experiments on the household-level electricity consumption of people in California, Michigan, and Texas. The results demonstrate the significant effectiveness of the ILB program in assisting communities during emergency situations.
\end{itemize}

\section{Related Work}\label{Sec: related work}

\subsection{Demand Response Programs}

A critical advantage of DR programs lies in their potential to substantially mitigate peak demand - a pressing issue in electricity systems that necessitates significant investment in underutilized infrastructure. Violette et al. \cite{violette2016} have demonstrated the potential of DR programs to curtail peak demand by up to $15\%$. Moreover, DR programs have been effective in enhancing the reliability of the electricity system, with comprehensive reviews like the one by Albadi and El-Saadany~\cite{albadi2008} elucidating a reduction in the frequency and duration of blackouts.

Existing DR programs fall into two broad categories: price-based programs (PBP) and incentive-based programs (IBP). PBP programs typically revolve around dynamic pricing rates, with the objective of reducing consumption during peak times. Notable approaches in this category include Time of Use~\cite{torriti2012}, where the rate changes based on the time of day, and Real-Time Pricing~\cite{bohn2002}, where the price fluctuates according to the market price of electricity. The second category, i.e. IBP programs, often reward participants with financial benefits either for their performance or for their participation. Direct Load Control (DLC) is one such program that allows the utility to control certain appliances within a consumer's home, such as air conditioning, in exchange for a financial incentive~\cite{spees2013}. Similarly, the Demand Bidding/Buyback program provides consumers the opportunity to bid for payments in exchange for their desired load reductions~\cite{cappers2013}.

\subsection{Time Series Forecasting}

Time-series methods can be grouped into two broad categories: univariate time series techniques and multivariate time series techniques. Univariate time series methods focus on analyzing individual observations sequentially without considering the correlations between different time series. ARIMA methods \cite{liu2016online}, for instance, assume linearity, where the prediction is a weighted linear sum of past observations. 
On the other hand, multivariate time series techniques consider and model the interactions and co-movements among a group of time series variables~\cite{shaham2023holistic,hajisafi2023learning}. Wan et al.~\cite{wan2019multivariate} propose an encoder-encoder model based on attention mechanism to capture correlations. The proposed end to end approach include bi-directional long short-term memory networks (Bi-LSTM) layers as the encoder network to adaptively learning long-term dependency and hidden correlation features of multivariate temporal data. The authors in \cite{shih2019temporal} use filters to extract temporal patterns that remain consistent over time. Then, apply attention mechanism to select pertinent time series and utilizes their frequency domain information for multivariate prediction.

\newcommand{\rvec}{\mathrm {\mathbf {r}}} 
\begingroup
\begin{table}
\caption {Summary of Notations.} 
\centering
\begin{tabular}{>{\arraybackslash}m{2.7cm} >{\arraybackslash}m{5cm} }
\hline\hline
  Symbol  & Description\\    \hline
  $n,\, m$ & Number of households and neighborhoods\\
  $\mathcal{U} = \{ u_1,...,u_n \}$ & Set of users\\
  $X_d^{\text{supply}},\, X_d^{\text{demand}}$ & Supply and demand power on day $d$\\
  $X_d^u$ & Power consumption of user $u$ on day $d$\\
  $k$ & Number of features\\
  $x^u$ & User $u$ power consumption time series\\
   $\mathcal{D},\mathcal{D}'$& Set of all days and emergency days\\
   $r_{\text{baseline}}^u,\, r_{\text{emergency}}^u$ & Baseline \& emergency price rate on ILB participants\\
   $I^u$ & Incentive for household $u$\\
   $e^u$ & PE of household $u$\\
   $A_{real},\, A_{est}$ & Real and estimated similarity matrices\\
\hline\hline
\end{tabular}
\label{tab:table1}
\end{table}
\endgroup

\section{Preliminaries}

\subsection{Notation}

Consider a map that contains $n$ households $\mathcal{U} = \{ u_1,...,u_n \}$, which are distributed among $m$ non-overlapping neighborhoods $\mathcal{N} = \{ N_1,..., N_m\}$. Each neighborhood $N_i$ contains a population of size $|N_i|$. We represent the hourly electricity consumption of household $u\in \mathcal{U}$ using a time series $x^u\in \mathbb{R}^{1\times T}$, and we denote the set of all time series by $X = (x^1,...x^n)\in \mathbb{R}^{n\times T}$. Additionally, we model the total power consumption of user $u$ on day $i$ as $X_i^u$, and the set of all days in a billing cycle by $\mathcal{D}$. Table~\ref{tab:table1} summarizes the important notations used throughout the manuscript.


\subsection{Message Passing Framework}

Most GNNs use message-passing and aggregation to learn improved node representations, or so-called embeddings, of the graph. At propagation step $i$, the embeddings of node $v$ is derived by:
\begin{equation}
    H_v^{i} = f(\text{aggr}(H_u^{i-1}|u\in N_1(v))).
\end{equation}
In the above formulation, $H_v^i$ represents the $i$-th set of features for the nodes, $f(.)$ is the function used to transform the embeddings between propagation steps, $N_1(.)$ retrieves the 1-hop embeddings of a node, and $\text{aggr}$ combines the embeddings of its 1-hop neighbors. For instance, in GCN, node aggregation and message-passing are expressed as:

\begin{equation}
H_v^{i} = \sigma(W \sum_{u\in N_1(v)}  \dfrac{1}{\sqrt{\hat{d}_u\hat{d}_v}} H_u^{i-1} ),
\end{equation}

where $\sigma$ is the activation function, $W$ is a matrix of learnable weights, $\hat{d}_v$ is the degree of node $v$. The propagation length of the message-passing framework is commonly limited to avoid over-smoothing.

\section{Proposed Scheme}
In this section, we present our proposed emergency DR program. 

\subsection{Program Overview}

With the rising trend of electricity consumption, combined with the limited capacity of supply, electricity disruptions in household networks are becoming inevitable. Suppose the utility provides $X_i^{\text{supply}}$ kWh of power on a particular day $i$, but the electricity demand is $X_i^{\text{demand}}$ kWh, where $X_i^{\text{supply}}< X_i^{\text{demand}}$. This necessitates a strategy to balance the supply and demand. Currently, there are two extreme strategies in place: (I) implementing the same level of power outage for all users by reducing each household's electricity by $( X_i^{\text{demand}} - X_i^{\text{supply}})/n$, and (II) focusing on only $l$ households and cutting their power for a longer duration, while ensuring uninterrupted power supply for the others. The second strategy results in an outage of $( X_i^{\text{demand}} - X_i^{\text{supply}})/l$ in this group of households.

The sudden and unexpected power cuts can disrupt the daily life of households, regardless of their efforts to manage the power demand and response. In order to tackle this issue, we propose a price-driven demand response program named the Incentive-driven Load Balancer (ILB), which prioritizes the user's preferences. The ILB program enables households to voluntarily participate in a program that offers them financial incentives upfront, in exchange for paying higher rates during a few unanticipated days. The households are informed of such occurrences a day before and encouraged to reduce their electricity consumption during those times. For those who do not opt in to the program, higher rates will be charged throughout the period to cover the program's expenses. The formal definition of the ILB program is provided in Definition~\ref{ILB}.

\begin{defn}\label{ILB}
    (\emph{Incentive-Driven Load Balancer}). 
     Under this strategy, households are allowed to voluntarily participate in a program, where they agree to pay higher rates for a number of unplanned days during the upcoming billing cycle, which they will be notified of only one day in advance. In return for their participation, they receive a monetary incentive at the start of the program. For the remaining duration of the billing cycle, the standard rates will be applied. To cover the cost of the incentives, the electricity rate for all other households is increased.
\end{defn}

The aim of ILB to incentivize flexible users who opt-in to reduce their power usage during emergency days, rather than enforcing power outages to balance demand and supply, or force higher prices on all users. This is achieved by charging the opt-in users higher rates, encouraging them to adjust their consumption behavior. 
The incentives provided to the users needs to be carefully determined --- if it is too low, then not enough customers will accept the offer resulting in an insufficient reduction of the demand during the emergency days; if too high, it may potentially result in too high a number of households signing up for the incentive making the program too expensive, raising the burden on non-participating households beyond acceptable levels. 

To quantify the utility of the proposed program, we propose the following two indicators. The first utility function aims to reveal if offers are successful in attracting customers.

\begin{defn}
    (\emph{Utility: Acceptance Rate}). The acceptance rate indicates the percentage of households who agree to participate in the program after receiving an offer. 
    \begin{equation}
    \text{Acceptance rate} = \frac{\#\, \text{accepted offers}}{\#\,\text{offers made}} \times 100.
    \end{equation}
\end{defn}

The second utility focuses on the amount of reduction made in consumption given the incentives formulated in Definition~\ref{definition: company}.

\begin{defn}\label{definition: company}
    (\emph{Utility: Responsiveness Cost}). Let $\mathcal{U}'$ denote the set of users who voluntarily participate in the program, and $\mathcal{D}' = { d_i,...,d_j}$ denote the set of emergency days that occurred during the billing cycle. The responsiveness cost of ILB is defined as
    \begin{equation}
    \text{Responsiveness Cost} =  (\sum_{u\in \mathcal{U}'} I^u ) / (\sum_{d\in D'} \sum_{u\in \mathcal{U}'} \Delta X_d^u).
    \end{equation}
    In the above formulation, $\Delta X_d^u = \Bar{X}_i^u - X_i^u$ represents the variation in the consumption of user $u$ on day $d$. Here, the consumption of user $u$ is represented by $X_d^u$, and the adjusted consumption due to their participation is represented by $\Bar{X}_d^u$. The symbol $I^u$ refers to the incentive provided to user $u$ at the beginning.
\end{defn}

The rationale behind the responsiveness utility function is to determine the amount of power consumption reduction that can be achieved for a given amount of incentives to participants. The utility company uses the following optimization formulation to balance supply and demand while minimizing incentive expenditure. The responsiveness cost is a metric that measures how well this objective is met in practice.

\begin{equation}
\begin{aligned}
& \text{Minimize}
& &  \sum_{u\in \mathcal{U}'} I^u    \\
& \text{subject to}
& &  X_d^{\text{demand}} - X_d^{\text{supply}} \leq  \sum_{u\in \mathcal{U}'} \Delta X_d^u,\,\,\, \,\,\, \forall \, d\in \mathcal{D}' \\
\end{aligned}
\end{equation}

\subsection{Pricing}

As utility functions outlined in the previous section reveal, ILB requires a diligent selection of participants and careful pricing of incentives. In this subsection, we address the latter and explain how incentives are calculated, and in Section~\ref{Implementation Framework}, the selection process of applicants is illustrated. Recall that $\mathcal{D}'\subset \mathcal{D}$ are the emergency days.
For any $u\in \mathcal{U}'$, let us denote the price rate for emergency days ($d\in \mathcal{D}'$) by $r_{\text{emergency}}^u$ and their regular rate for $d\in D \textbackslash  D'$ by $r_{\text{baseline}}^u$. Thus, by participating in the program the cost of user $u$ will be modified from their baseline cost,

\begin{equation}
    Cost_{\text{baseline}} (u) = \sum_{d\in D} X_d^u \times r_{\text{baseline}^u},
\end{equation}

\noindent to the modified cost of,

\begin{equation}
    Cost_{\text{ilb}} (u) = \sum_{d\in D \textbackslash  D'} X_d^u \times r_{\text{baseline}}^u + \sum_{d\in D'} \Bar{X}_d^u \times r_{\text{emergency}}^u - I^u.
\end{equation}
Once it comes to consideration, the household should understandably expect
\begin{equation}\label{Equation: inequality}
 Cost_{\text{baseline}} (u) \geq Cost_{\text{ilb}} (u).
\end{equation}
Otherwise, the offer will not be beneficial for the user. The households who are not part of the program will be charged by an extra charge of $r_{\text{extra}}$ and pay the modified rate of,

\begin{equation}
    r_{others} = r_{\text{baseline}} + r_{\text{extra}}.
\end{equation}


The rate hike for users not included in the program is derived based on the incentives provided in ILB, calculated as,
\begin{equation}
     r_{\text{extra}} =  (\sum_{u\in \mathcal{U}' } I^u)/(\sum_{u\in  \mathcal{U}\textbackslash \mathcal{U}'} \sum_{d\in D} X_d^u).
\end{equation}

\subsection{Rate Hikes}

A critical factor in the above formulation is the rate hikes on emergency days. This factor directly influences the amount of demand response on such days. We derive this rate based on the {\em price elasticity of demand for electricity}.

\begin{defn}
    (\emph{Elasticity of Demand}). The elasticity of demand refers to the degree to which demand responds to a change in an economic factor.
\end{defn}

The appropriate value of $r_{\text{emergency}}$ must be chosen such that the increase in rates results in a change in consumers' demand, which can be measured by the price elasticity of demand. The price elasticity of demand is derived by,

\begin{equation}
    PE = \dfrac{\%\, \text{Change in Quantity}}{\%\, \text{Change in price}}.
\end{equation}

The equation indicates the percentage of change in demanded quantity for a given percentage of change in price. For example, if the price of electricity increases by $10\%$, and the quantity of electricity demanded decreases by $5\%$, the price elasticity of demand for electricity can be calculated as:

\begin{equation}
    PE= (-5\% / 10\%) = -0.5
\end{equation}
The PE for electricity has been measured and estimated in many countries in the world including the US.  The factor is commonly considered for the short term and long term. The average PE for electricity on the state level is estimated to be $-0.1$ in the short-run and $-1.0$ in the long-run~\cite{burke2018price,miller2016sensitivity}.


It is crucial to note that the numbers mentioned earlier represent average statistics for households. However, on an individual household level, the PE for electricity can vary greatly, which can impact the necessary amount of incentive required for them to accept an offer. Let us denote the PE for household $u$ by $e^u$. Hence,  given the goal of ILB that every household in the program reduces its consumption by $i\%$, the percentage of change in price for the household is calculated by

\begin{equation}
\%\, \text{Change in price} = i/ e^u.
\end{equation}
Once the percentage of change in price is calculated, the rates on emergency days are calculated by
\begin{equation}
    r_{\text{emergency}}^u =(1+ \%\, \text{Change in price})\times  r_{\text{baseline}}^u.
\end{equation}

Armed with this knowledge, applying inequality in the Equation~\ref{Equation: inequality}, the minimum incentive values can be derived as,

\begin{align}
& Cost_{\text{baseline}} (u) \geq  Cost_{\text{ilb}} (u)  \rightarrow\\
\begin{split}
    & \sum_{d\in D} X_d^u \times r_{\text{baseline}}^u \geq 
    \quad \sum_{d\in D \setminus D'} X_d^u \times r_{\text{baseline}}^u\\ 
    & \quad \quad  \quad  \quad  \quad  \quad  \quad  \quad  \quad  + \sum_{d\in D'} \Bar{X}_d^u \times r_{\text{emergency}}^u - I^u \rightarrow
\end{split}\\
     & I^u \geq  \sum_{d\in D'} \Bar{X}_d^u \times r_{\text{emergency}}^u - \sum_{d\in D'} X_d^u \times r_{\text{baseline}}^u.\label{Equation: make sense}
\end{align}

\begin{figure*}[t]
\includegraphics[scale=.38]{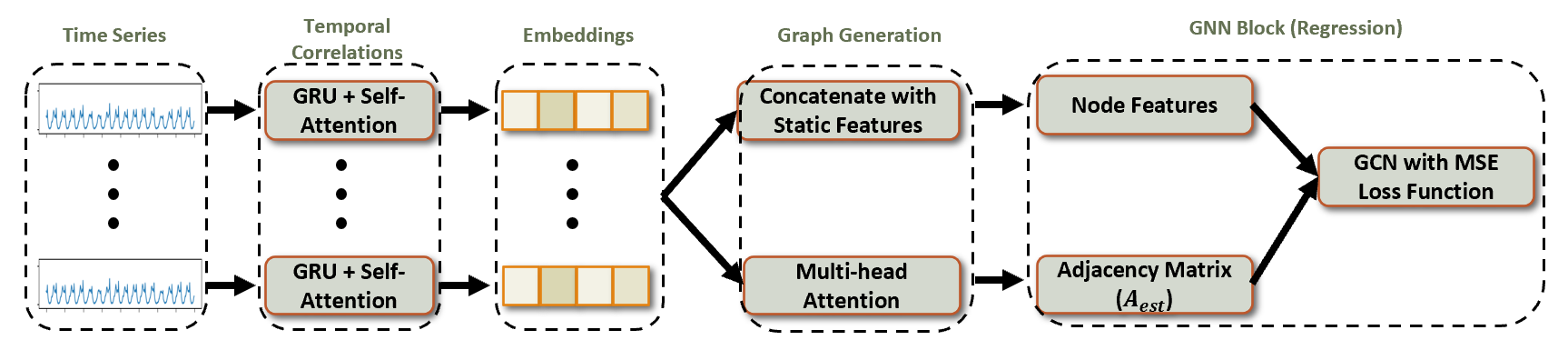}
\centering
\caption{Pattern Recognition Graph.}
\vspace{-10pt}
\label{Fig: pattern recognition graph}
\end{figure*}

\section{Implementation Framework}\label{Implementation Framework}

The success of the program relies heavily on the process of selecting candidates. Utilizing both a sequential and machine learning approach can aid in identifying when enough households have been efficiently selected, offered, and accepted the incentive. This is essential in ensuring that a satisfactory number of high-demand flexibility users have accepted the incentives and that the expected reduction in demand meets the shortfall on emergency days. Additionally, this method prevents over-recruitment, which could result in unnecessary program expenses.

Our proposed approach for choosing potential participants for the ILB program is based on two graph models: the pattern recognition GNN and the household selection GNN. The dynamic pattern recognition model is designed to generate a similarity matrix among users, which indicates the probability that a household would respond positively to an offer based on the response of its neighbors and other households. The household selection model, on the other hand, is a node classification model that aims to identify the candidates who are most likely to accept the offer.

\subsection{Pattern Recognition GNN}

The main objective of the pattern recognition model is to create an accurate similarity matrix that reflects the degree of similarity between two individuals based on their socio-economic status and electricity consumption pattern. This advance model helps us understand the likelihood of a household accepting an offer, given the responses of those already queried. The underlying assumption is that households with a high similarity score are likely to respond similarly to the offer. Although there may be some margin of error, a well-designed model can mitigate this issue to a reasonable extent. Our proposed model considers three factors: (I) socio-economic factors affecting demand response, (II) intra-series temporal correlations in time series, and (III) inter-series correlations captured by an attention mechanism that reveals pattern similarity. While the model is primarily designed to predict future demand, the inter-series attention matrix enables enhanced modeling of user similarity.


We use $A_{real}\in \mathbb{R}^{n\times n}$ and $A_{est}\in \mathbb{R}^{n\times n}$ to denote the actual and predicted probability matrices of accepting the offer, respectively. The element ($a_{ij}$) located in the $i$-th row and $j$-th column of these matrices represents the likelihood of household $u_i$ accepting the offer, given that household $u_j$ has already accepted the offer. An overview of the model is provided in Figure~\ref{Fig: pattern recognition graph} and the components of the model are elaborated layer by layer in the following.

\subsubsection{Intra-series Temporal Correlation Layer} The first layer of the model aims to capture the temporal correlation in each time series. For a given window $s$, the training data $X\in \mathbb{R}^{n\times s}$ would be input to the model. Each time series is processed by two essential components within this layer: a Gated Recurrent Unit (GRU) unit and a self-attention module.

The purpose of GRU units is to handle sequential data by temporal dependencies. GRU has been shown to utilize less memory and perform faster compared to Long Short Term Memory (LSTM). The self-attention components have been included after GRU units to enhance their performance further. The input data $X$ is fed into GRU units and self-attention components generating embeddings $C = (c_1,...,c_s)\in \mathbb{R}^{s\times M}$ where $M$ is the embedding size.

\subsubsection{Inter-series Correlation Layer}
Next, the generated embeddings are fed to a multi-head attention layer~\cite{vaswani2017attention}. The multi-head attention layer consists of multiple attention heads, each of which learns to attend to different parts of the input sequence.  This is the critical step where the correlation and similarity between households are captured. The ultimate output of this unit is the matrix $A_{\text{est}}$, or the so-called attention matrix, representing the pairwise correlations of households.

The input of the attention layer, as formulated in Equation~\ref{equation: multihead attention 1} consists of queries ($Q$), keys ($K$), and values ($V$), each of which is a sequence of vectors. To understand the similarity between time series, all inputs are set to the embedding matrix $C$ generated in the previous layer. The layer then applies multiple attention heads, each of which computes a weighted sum of the values using a query and key pair. The outputs from each attention head are concatenated and linearly transformed to produce the final output of the layer.

\begin{equation}\label{equation: multihead attention 1}
A_{est}=MultiHead(Q,K,V)=(head_1\mathbin\Vert…\mathbin\Vert head_j)W^O,
\end{equation}
where each head consists of, 

\begin{equation}
head_i=Attention(QW_i^Q,KW_i^K,VW_i^V),
\end{equation}
\begin{equation}
Attention\left(Q,\ K,\ V\right)=Softmax\left(\frac{QK^T}{\sqrt{M}}\right)V.
\end{equation}
In the above formulation, $W^O$, $W_i^Q$, $W_i^K$, and $W_i^V$ are weight matrices learned during the training.

\subsubsection{Socio-economic Factors}

Next, the embedding produced in the preceding layers will be combined with the socio-economic factors to serve as node features in the dynamic graph. It is important to note that our focus is on factors that have been proven to have a considerable effect on demand elasticity, not just on user consumption. For instance, even though the number of individuals in a household can impact consumption, it has not been identified as a significant coefficient of demand elasticity in some studies, as cited in~\cite{du2015residential}. We have summarized the key determinants in the following. (I) {\em Income}: Studies, such as the one by Brounen et al.\cite{brounen2012residential}, have demonstrated that the level of income has a notable influence on the elasticity of demand for electricity. (II) Demographic characteristics: Factors such as race and age significantly impact household electricity consumption\cite{filippini2004elasticities,du2015residential}. (III) Dwelling physical factors: Aspects including the type of building, its size, and its thermal and quality characteristics are linked to the amount of energy consumed by a household~\cite{tso2003study}. (IV) Climate: Temperature has been demonstrated to impact electricity consumption significantly~\cite{du2015residential}. (V) Living Area: The geographical location of households, including regional variables such as urban and rural areas, cities and counties, critically affect their elasticity of demand~\cite{du2015residential}. (VI) Awareness and Education: The level of awareness about policies and the education level of individuals can influence their electricity demand and compliance with ILB, as shown by Du et al.~\cite{du2015residential}, who found the coefficient representing the extent to which users understand the policy to be a significant factor in demand response.

Let us denote the socio-economic features for $i$-th household by $\hat{c}_i\in \mathbb{R}^{\bar{M}}$. The concatenation of embeddings from time series and socio-economic features leads to node features of the dynamic graph, mathematically represented by,

\begin{equation}\label{Equation: node features}
    c_i' = c_i || \hat{c}_i \in \mathbb{R}^{(M+\bar{M})},
\end{equation}
resulting in the final node feature matrix $C' = (c_1',...,c_{n}')\in \mathbb{R}^{n\times (M+\bar{M})}$.

\subsubsection{Dynamic Graph}

In the final stage of the pattern recognition graph, the dynamic GNN is formulated as $G(V,E)$, where $V$ represents the node set and $E$ represents the edge set. To create the graph, each household is assigned a node, and their node features are generated using the concatenated features derived from Equation~\ref{Equation: node features}. The attention matrix $A_{est}$ obtained from Equation~\ref{equation: multihead attention 1} is utilized as the edge weights of the graph. Subsequently, the GNN block is applied to the graph. The result of this process is the embeddings for the nodes, represented by

\begin{equation}
    C'' = F_{\text{GNN}}(G(V,E)) \in \mathbb{R}^ {n\times M"}.
\end{equation}


where $F_{\text{GNN}}$ denotes a custome GNN model. The graph uses the mean square error (MSE) as its objective loss function to predict demand. After the training process, the pattern recognition graph outputs the attention matrix $A_{\text{est}}$, which is utilized to represent the similarity between households. 


\begin{figure}[t]
\includegraphics[scale=.4]{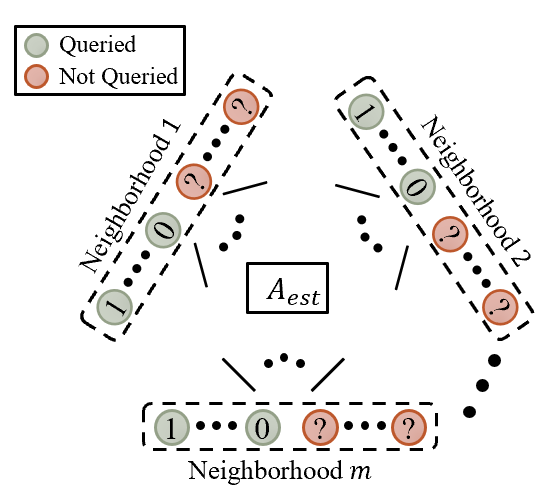}
\centering
\caption{Household Selection Graph.}
\vspace{-6pt}
\label{Fig: Household Selection Graph}
\end{figure}

\subsection{Household Selection GNN}
The purpose of the household selection graph is to efficiently select the users who have the highest likelihood of participating in the program. Therefore, the graph is a node classification GNN conducting the semi-supervised task. Let $G_{\text{HSG}}= (V_{\text{HSG}}, E_{\text{HSG}})$ denote the household selection graph. The similarity matrix generated based on the pattern recognition graph represents the weighted edges of the graph, i.e., $E_{\text{HSG}}= A_{\text{est}}$. At this stage all socio-economic and power consumption information have been incorporated in the similarity matrix. The set of nodes in the graph consist of a single node for each household, i.e., $V_{\text{HSG}}= \mathcal{U}$. The one-hot encoding method is used as node features of the graph. For example, if $4$ households exist in the network, their feature set would be $0001,\, 0010,\, 0100,\, 1000$. 

The goal of this graph is to efficiently label nodes indicating acceptance or rejection of an offer. Two steps are taken to conduct the labeling. First, Spectral Graph Analysis is conducted on the graph to cluster nodes based on the similarity. Such an approach allows for naive understanding of the node labels. Based on the spectral graph analysis small portion of households in each neighborhood are queried to understand the true labels. Then, those labels are used in the graph to conduct the semi-supervised graph labeling.

\subsubsection{Spectral Graph Analaysis}

Spectral clustering methods rely on the spectrum (eigenvalues) of the data's similarity matrix to reduce its dimensions before clustering it in a lower number of dimensions. The similarity matrix $A_{\text{est}}$ is given as input to the clustering. 

First, the method computes the normalized Laplacian of the graph as defined by the following equation:

\begin{equation}
L = I' - D''^{-1/2} A_{\text{est}} D''^{-1/2}.
\end{equation}

In this equation, $I'$ represents the identity matrix, while $D''$ is defined as $diag(d)$, where $d(i)$ denotes the degree of node $i$.

Secondly, it computes the first $k$ eigenvectors that correspond to the $k$ smallest eigenvalues of the Laplacian matrix. Once eigenvectors are derived, a new matrix with the $k$ eigenvectors is generated, where each row is treated as a set of features of the corresponding node in the graph. Finally, the nodes are clustered based on these features using $k$-means algorithms into $2$ clusters representing people who are likely to accept or reject the offer.

\subsubsection{Semi-Supervised Node classification}

Up to this point, there have been no offers made to determine real-world responses to offers. However, using spectral graph analysis, households have been grouped into two clusters and assigned a label. It's important to note that the labels based on these clusters don't necessarily indicate which group is more likely to accept the offers. The ultimate goal at this stage is to use the perceptions received from clustering to survey a small portion of individuals from each of the $m$ neighborhoods, and use this data alongside the adjacency matrix in a semi-supervised approach to identify individuals who are likely to accept the offers.

To ensure that all parts of the graph are properly discovered and fairly treated, it's crucial to diversify the initial queries in all neighborhoods. For this reason, in each neighborhood, $5\%$ of users from each cluster generated by spectral graph analysis are selected to be offered incentives. Querying this sample of users leads to real labels determined for $10\%$ of the population.
With this information in hand, a node classification GNN model is applied on top of $G_{\text{HSG}}= (V_{\text{HSG}}, E_{\text{HSG}})$ to classify whether other users will accept the offer or not. The resulting labels are used to make offers.

\section{Experimental Evaluation}

\subsection{Datasets}


We conduct our experiments by combining multiple datasets from various sources, taking into consideration significant aspects such as geographical diversity, the time series of household electricity consumption, and socioeconomic factors.

{\em Household-level Electricity Time-Series.} The privacy concerns regarding household-level electricity consumption have limited the availability of publicly accessible datasets, with current datasets containing no more than $50$ households. A comprehensive overview of these datasets can be found in~\cite{babaei2015study}. Nonetheless, the work in~\cite{thorve2023high} has tackled this issue by creating a synthetic dataset that covers households throughout the United States. This dataset functions as a digital twin of a residential energy-use dataset within the residential sector. For our experiments, we use the time-series data at the household level from this dataset.

{\em Education and Awareness.} To indicate the level of awareness, the average ACT scores are used, which are obtained from the EdGap dataset~\cite{edgap}. This dataset contains socioeconomic features of high school students across the United States. The geospatial coordinates of schools are derived by linking their identification number to data provided by the National Center for Education Statistics~\cite{education}, and then the average ACT scores are calculated. Additionally, the percentage of people who have completed some college education provided by~\cite{usda} is used to further expatiate on the level of awareness. 

{\em Median Household Income and Unemployment Percentage.} The county-level information for the median household income as well as the unemployment percentage in 2014 is extracted from the US Department of Agriculture website~\cite{usda} and used as static features for counties. 

{\em Climate.} Two key attributes provided by the National Centers for Environmental Information~\cite{NCEI} are used as indicators of climate in counties: average temperature and precipitation amount. 

\begin{figure*}[t]
	\subfloat[California\label{figure: heatmap s1}]{  
	\includegraphics[ scale = .21]{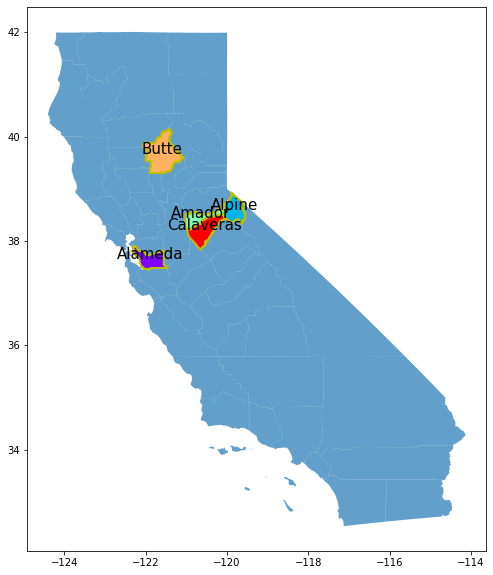}  
	}
	\hfill
	\subfloat[Michigan\label{figure: heatmap s2}]{%
	\includegraphics[ scale = .21]{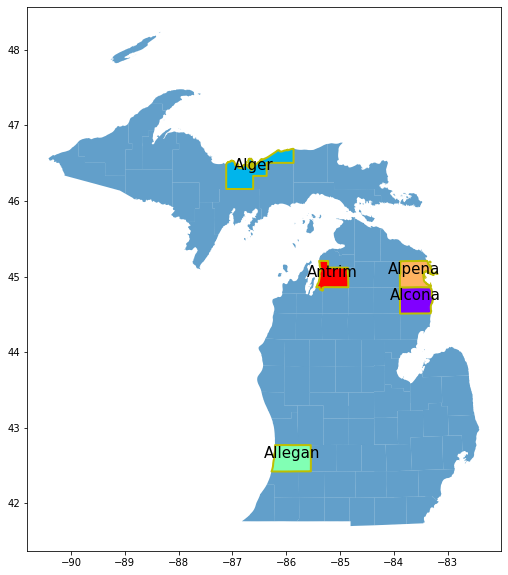}
	}
	\hfill
	\subfloat[ Texas\label{figure: heatmap s3}]{%
	\includegraphics[ scale = .21]{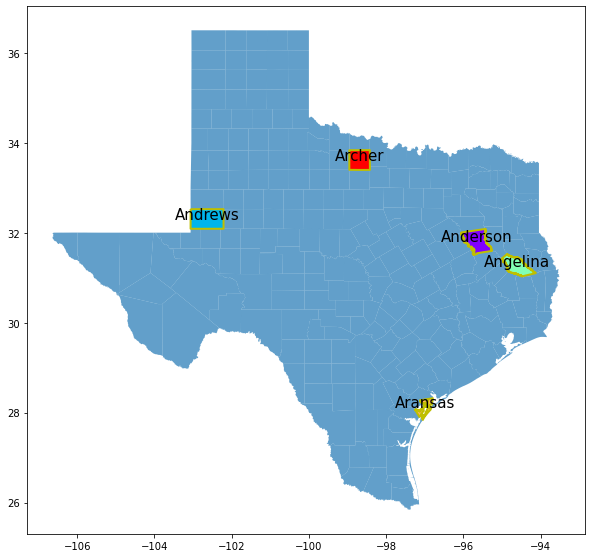}
	}
    \newline
	\subfloat[California\label{figure: heatmap s4}]{%
	\includegraphics[scale=.28]{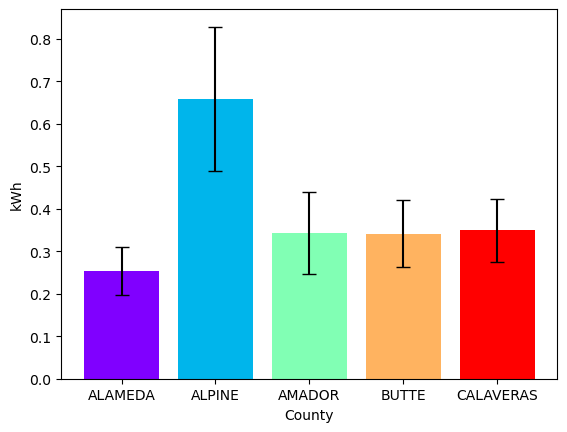}
	}
	\hfill
	\subfloat[Michigan\label{figure: heatmap s5}]{%
	\includegraphics[scale=.28]{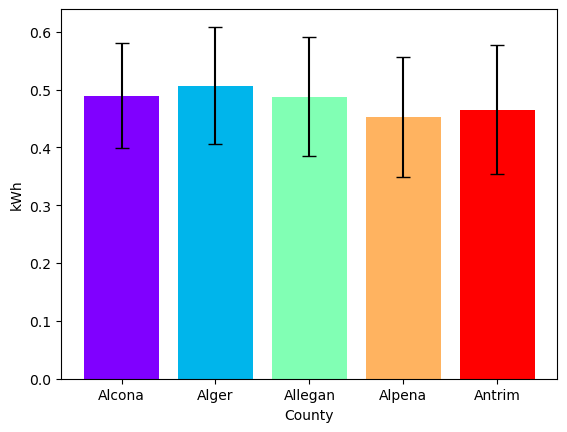}
	}
	\hfill
	\subfloat[ Texas\label{figure: heatmap s6}]{%
	\includegraphics[scale=.28]{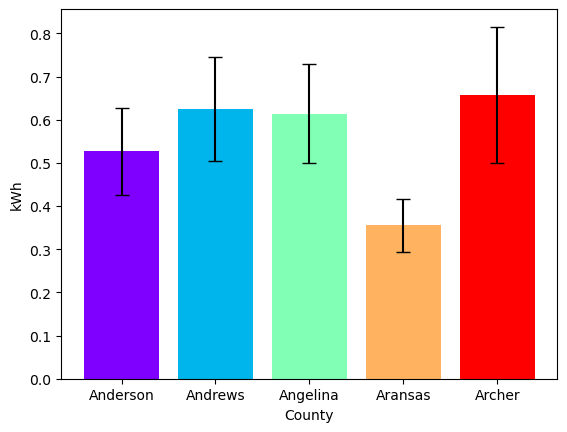} 
	}
	\vspace{-5pt}
	\caption{Counties used for the purpose of experiments and their corresponding average hourly consumption. The standard deviation of consumption is shown as error bars. }
    \label{figure: stat}
	\vspace{-6pt}
\end{figure*}

{\em Geographic Diversity.} Three states of California (CA), Texas (TX), and Michigan (MI) are selected for the purpose of experiments, including the first five counties based on codes published by the U.S. Census Bureau~\cite{bureau2022}. 
The geospatial location of counties used in experiments and their corresponding statistics is provided in Figure~\ref{figure: stat}.

\subsection{Experimental Setup}

{\em Overview.} The experiments are divided into two parts, each looking at performance before and after applying the framework. Subsection~\ref{subsection: Performance Analysis of ILB} specifically looks at how well the ILB performs before the selection framework is applied, and investigates how the program affects responsiveness cost and how much it can reduce the total consumption given a certain amount of incentive. Subsection~\ref{subsection: Evaluating the Integrated Framework and ILB} focuses on the performance after the framework has been applied in which factors such as responsiveness cost, the effects of rate increases on those not participating in the program, and noise analysis in the selection process are thoroughly evaluated.

{\em Hyper-parameters Setting.} The household electricity time series is used on an hourly basis between September and December of $2014$. The number of households participating in the program in each county is $50$, adding up to $250$ households in each state. The dataset was separated into three parts: training, evaluation, and testing sets, with a ratio of 7:2:1 respectively. Z-normalization was applied to normalize the input data, and training was conducted using the RMSProp optimizer with a learning rate of 3e-4. Training took place over $100$ epochs, with a batch size of $32$. The number of emergency days per month is set to three days. The number of participants in the experiment is set to be one-quarter of the population being considered and the default incentive provided to participants is $100$ dollars unless stated otherwise. The electricity PE for households is randomly selected based on a Gaussian distribution with a mean of $-0.25$ and a standard deviation of $0.1$.

{\em Hardware and Software Setup.} Our experiments were performed on a cluster node equipped with an $18$-core Intel i9-9980XE CPU, $125$ GB of memory, and two $11$ GB NVIDIA GeForce RTX 2080 Ti GPUs. Furthermore, all neural network models are implemented based on PyTorch version 1.13.0 with CUDA 11.7 using  Python version 3.10.8.

\subsection{Performance Analysis of ILB}\label{subsection: Performance Analysis of ILB}

In this subsection, we focus on the ILB program's performance independent of the framework. Initially, we evaluate the ILB's efficiency in terms of responsiveness cost, and subsequently, the program's effectiveness in total demand reduction of electricity.

\begin{figure*}[tbh]
	\subfloat[California\label{figure: disparity proof s4}]{%
	\includegraphics[scale = 0.35]{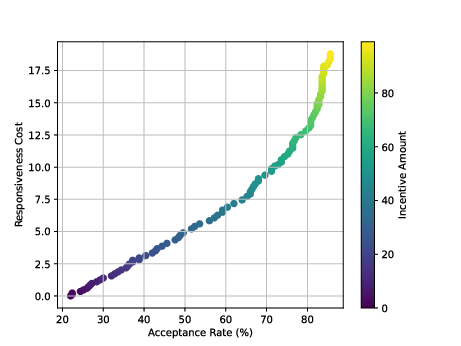}
	}
	\hfill
	\subfloat[Michigan\label{figure: disparity proof s5}]{%
	\includegraphics[scale = 0.35]{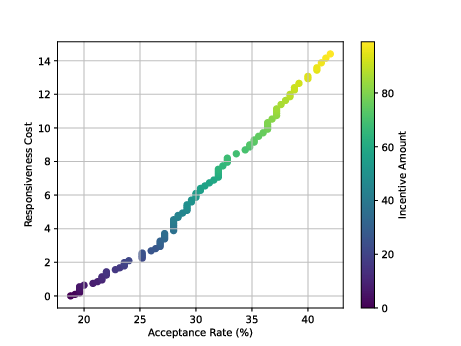}
	}
	\hfill
	\subfloat[Texas\label{figure: disparity proof s6}]{%
	\includegraphics[scale = 0.35]{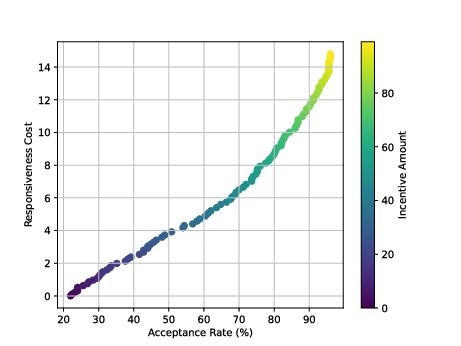}
	}
        \hfill
	\vspace{-6pt}
	\caption{Evaluation of total responsiveness cost.}
    \label{figure: responsiveness cost}
	\vspace{-6pt}
\end{figure*}

\subsubsection{Responsiveness Cost}\label{section: Responsiveness Cost}


Figure~\ref{figure: responsiveness cost} illustrates the results for the ILB program's responsiveness cost. Recall that responsiveness reflects the total incentive amount offered in relation to the observed reduction, thus a lower cost indicates better performance. The figure showcases the performance across three states: California, Michigan, and Texas. In each subfigure, given a specified incentive (highlighted with a bar) offered to the entire community, each individual assesses the benefit of participating in the program. This assessment is conducted using Equation~\ref{Equation: make sense}, which allows us to derive the community's acceptance rate and represent it on the $x$-axis, while the corresponding responsiveness cost is plotted on the $y$-axis. 
The figure reveals that an increase in the incentive amount expectedly raises the acceptance rate, but it also results in a higher responsiveness cost. The rate of increase for responsiveness cost tends to be lower for smaller incentive amounts and grows for larger values.




\subsubsection{Total Demand Reduction}

\begin{figure*}[tbh]
	\subfloat[California\label{figure: disparity proof s4}]{%
	\includegraphics[scale = 0.33]{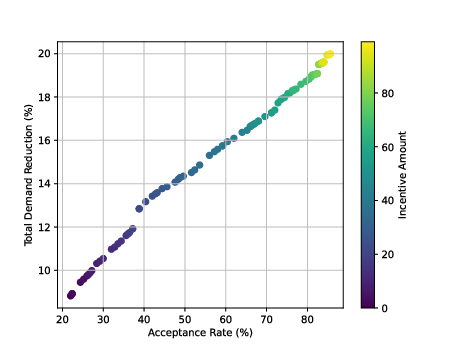}
	}
	\hfill
	\subfloat[Michigan\label{figure: disparity proof s5}]{%
	\includegraphics[scale = 0.33]{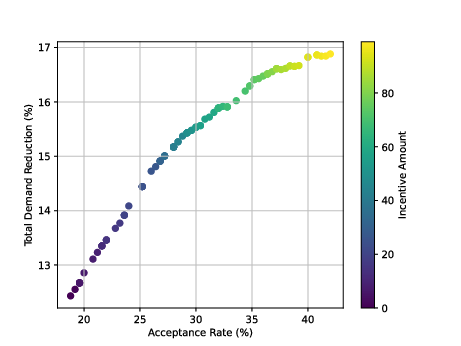}
	}
	\hfill
	\subfloat[Texas\label{figure: disparity proof s6}]{%
	\includegraphics[scale = 0.33]{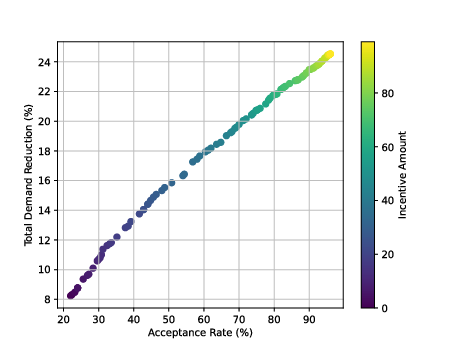}
	}
	\vspace{-6pt}
	\caption{Evaluation of total demand reduction.}
    \label{figure: total demand reduction}
	\vspace{-6pt}
\end{figure*}

Figure~\ref{figure: total demand reduction} shows the assessment of the total percentage reduction in consumption on emergency days. The structure of subfigures is analogous to Figure~\ref{figure: responsiveness cost}, but they illustrate the impact on the total consumption reduction instead of responsiveness cost on the $y$-axis. An increasing trend is observed across all three states indicating that a rise in the incentive amount and acceptance rate leads to a higher reduction in consumption on emergency days, enhancing the agility and performance of the electricity network. This pattern hints at a crucial trade-off between responsiveness cost and overall demand reduction, which becomes increasingly apparent as the incentives and participant count increase.

\subsection{Evaluating the Integrated Framework and ILB} \label{subsection: Evaluating the Integrated Framework and ILB}

Unlike the previous subsection, where offers were made to every household, this subsection evaluates performance based on a proposed framework that incorporates socio-economic factors and other determinants affecting the elasticity of demand for electricity when selecting participants.

\begin{figure*}[tbh]
	\subfloat[California\label{figure: disparity proof s4}]{%
	\includegraphics[scale = 0.31]{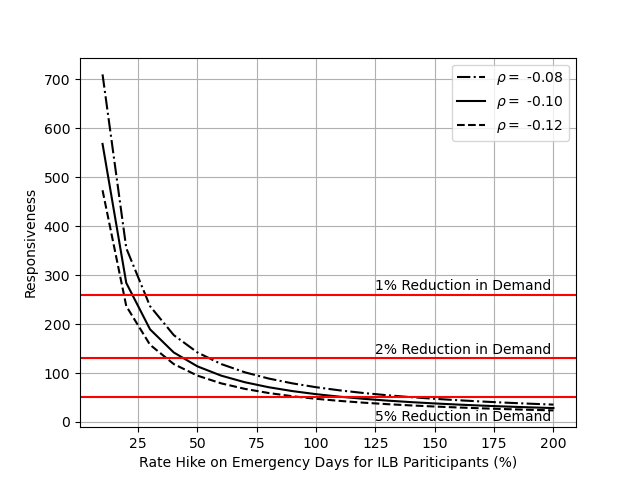}
	}
	\hfill
	\subfloat[Michigan\label{figure: disparity proof s5}]{%
	\includegraphics[scale = 0.31]{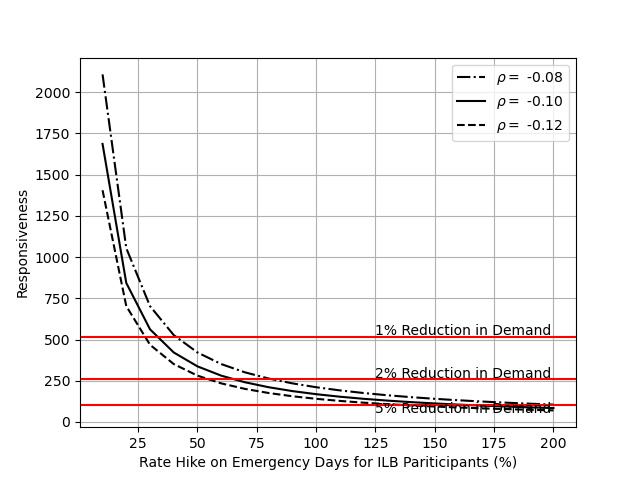}
	}
	\hfill
	\subfloat[Texas\label{figure: disparity proof s6}]{%
	\includegraphics[scale = 0.31]{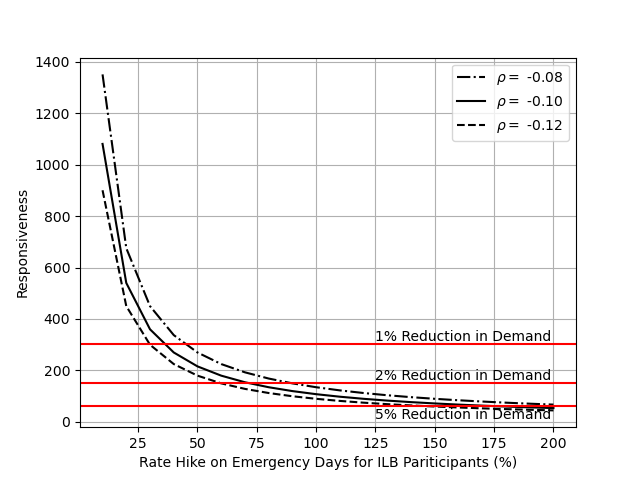}
	}
	\vspace{-6pt}
	\caption{Responsiveness performance of ILB.}
    \label{figure: Responsiveness Optimistic}
	\vspace{-7pt}
\end{figure*}

\begin{figure*}[tbh]
	\subfloat[California\label{figure: disparity proof s4}]{%
	\includegraphics[scale = 0.31]{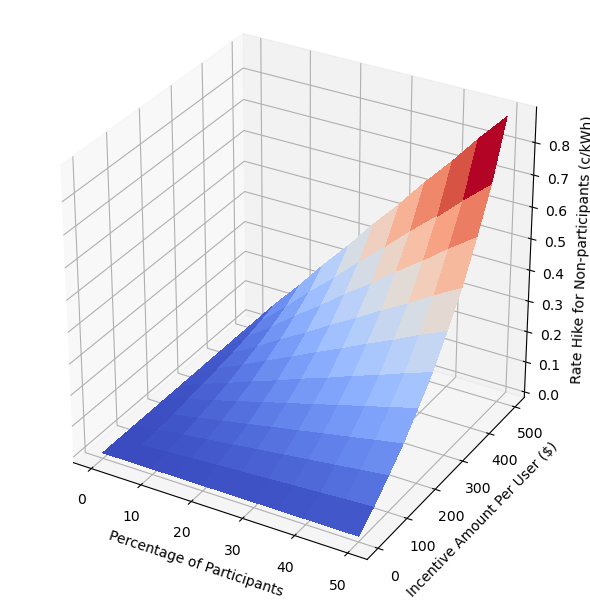}
	}
	\hfill
	\subfloat[Michigan\label{figure: disparity proof s5}]{%
	\includegraphics[scale = 0.31]{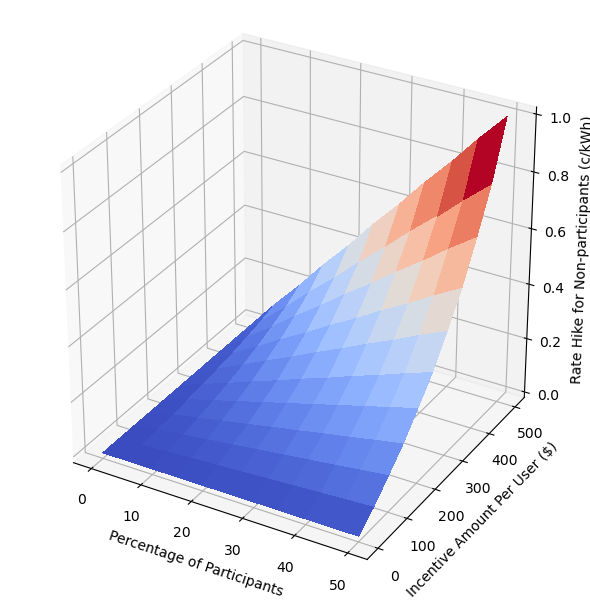}
	}
	\hfill
	\subfloat[Texas\label{figure: disparity proof s6}]{%
	\includegraphics[scale = 0.31]{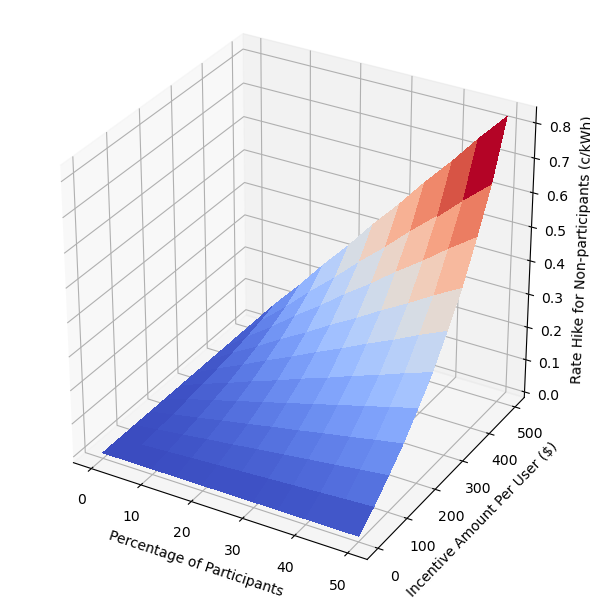}
	}
	\vspace{-6pt}
 \caption{Rate hike on nonparticipants in ILB.}
     \label{Fig: rate hike on nonparticipants}
	\vspace{-7pt}
\end{figure*}

\subsubsection{Responsiveness Cost}



The performance evaluation in terms of responsiveness cost is depicted in Figure~\ref{figure: Responsiveness Optimistic}. The $x$-axis represents the reduction in consumption by ILB program participants during emergency days, while the $y$-axis corresponds to the responsiveness cost. A quarter of households with the highest scores based on our framework are selected to participate in the program. As expected, once the amount of reduction by participants increases, a considerable decrease in responsiveness is observed in all three states. The horizontal red lines in the figure indicate the total reduction in demand for the whole community. The intersection of the curve and the horizontal line shows when the reduction in demand by ILB users accounts for a certain amount of total reduction in demand specified by the red line. The figures demonstrate that even small reductions of $10$ to $20$ percent by participants can lead to a significant reduction of approximately $5$ to $10$ percent in total demand. When the reduction in ILB participants is $50\%$, the total reduction accounts for $20$ to $25$ percent. Therefore, the proposed approach is a viable approach to effectively manage supply and demand during emergencies and prevent severe outages across states and counties.


\subsubsection{Rate Hikes on Non-participants}

Figure~\ref{Fig: rate hike on nonparticipants} shows the impact of the ILB program on non-participants. It displays the rate hikes in $c/kWh$ for non-participants at different percentages of ILB participation and corresponding incentives paid. Consistent with previous experiments, the PE for electricity is considered, and participants with the highest scores based on our framework are selected for the program. As it can be seen in the figure, increasing the percentage of participants and the amount of incentives paid leads to a higher cost for non-participants. However, the figure illustrates that in all three states, with a billing period of one month, by keeping the incentive amount within the range of $100$ to $200$ dollars, it is possible to only sacrifice a minimal cost on hourly consumption of non-participants.


\subsubsection{Candidate Selection and Noise Analysis}

In Table~\ref{table: noise analysis}, the proposed framework's effectiveness for selecting candidates to participate in ILB is presented alongside noise analysis on the similarity matrix. To introduce inaccuracies in the attention matrix, a uniform random noise is added to each entry of the matrix, following a distribution of $Uniform(0,b)$ where $b$ is set to three values: $25\%, 50\%$ and $75\%$ of the average value of the attention matrix. The results demonstrate that when the attention matrix is accurate, the model's accuracy is approximately $90\%$. However, as inaccuracies increase, the model's performance consistently declines across all three states. However, it is notable that even when the amount of inaccuracy is large, it does not severely impact the performance of the final model and the accuracy remains in an acceptable range. 


\begin{table}[t]
\caption{Accuracy of ILB candidate selection framework.}
\vspace{-6pt}
\label{table: noise analysis}
\begin{tabular}{|c|ccc|}
\hline
\textbf{Percentage of Noise } & \multicolumn{3}{c|}{\textbf{Accuracy}}                                  \\ \cline{2-4} 
                                              & \multicolumn{1}{c|}{California} & \multicolumn{1}{c|}{Texas} & Michigan \\ \hline
0 \%                                          &   \multicolumn{1}{c|}{88}      &   \multicolumn{1}{c|}{90}    & \multicolumn{1}{c|}{90}   \\ \hline
25 \%                                         &    \multicolumn{1}{c|}{84}     &   \multicolumn{1}{c|}{86}   &  \multicolumn{1}{c|}{87} \\ \hline
50 \%                                         &      \multicolumn{1}{c|}{81}   &    \multicolumn{1}{c|}{79}  &  \multicolumn{1}{c|}{86}  \\ \hline
75 \%                                         &    \multicolumn{1}{c|}{73}        &    \multicolumn{1}{c|}{78}  &  \multicolumn{1}{c|}{84}  \\ \hline
\end{tabular}
\vspace{-10pt}
\end{table}

\section{conclusion}
In conclusion, this paper introduced ILB, a novel program specifically designed for efficient demand management and response during crisis events. The ILB program encourages flexible households with the potential to lower their energy demand through incentives, thus facilitating effective demand reduction and preparing them for unforeseen circumstances. We also developed a two-step machine learning framework for the selection of participants. This framework, which incorporates a graph-based method to pinpoint households capable of readily modifying their energy usage, leverages two GNNs - one for recognizing patterns and another for household selection. Our comprehensive experiments on household-level electricity consumption across CA, MI, and TX provided compelling evidence of the ILB program's significant capacity to aid communities during emergency situations.

\bibliographystyle{IEEEtran}
\bibliography{sample}

\balance


\end{document}